\documentclass[journal]{IEEEtran}

\ifCLASSINFOpdf
\else
\fi

\usepackage[absolute]{textpos}
\usepackage[margin=0.674in,top=1in]{geometry}
\usepackage{algorithm}
\usepackage{algpseudocode}
\usepackage{amsmath, amssymb, bm, cite, epsfig}
\usepackage{epstopdf}
\usepackage{graphicx}
\usepackage{array}
\usepackage{multirow}

\usepackage{amsfonts}
\usepackage{tabularx} 
\usepackage{tabu}
\usepackage{pbox}
\usepackage{makecell}

\usepackage{booktabs}
\usepackage{ctable}
\usepackage{xcolor,colortbl}
\definecolor{Gray}{gray}{0.85}
\definecolor{LightCyan}{rgb}{0.8,1,1}

\def\beq{\begin{equation}}
\def\eeq{\end{equation}}
\def\beqa{\begin{eqnarray}}
\def\eeqa{\end{eqnarray}}
\def\beqan{\begin{eqnarray*}}
\def\eeqan{\end{eqnarray*}}

\def\Q{{\mathbb{Q}}}

\def\argmax{\mathop{\mathrm{arg\,max}}}

\def\tm1{t\! - \! 1}
\def\tp1{t\! + \! 1}

\def\log{\mathrm{log}}

\def\BS{\mathrm{BS}}
\def\MS{\mathrm{MS}}
\def\D{\mathrm{D}}
\def\Q{\mathrm{Q}}

\def\vec{\mathrm{vec}}
\def\RF{\mathrm{RF}}

\def\BB{\mathrm{BB}}

\def\for{\mathrm{for}}
\def\otherwise{\mathrm{otherwise}}
\def\can{\mathrm{can}}
\def\argmax{\mathrm{argmax}}
\def\Tr{\mathrm{Tr}}
\def\and{\mathrm{and}}

\begin{document}
	\begin{textblock}{18.7}(1.5,0.5)
		\centering
		\noindent\Large S. Sun and T. S. Rappaport, "Millimeter wave MIMO channel estimation based on adaptive compressed sensing," \textit{2017 IEEE International Conference on Communications Workshop (ICCW)}, Paris, May 2017.
	\end{textblock}
\pagenumbering{gobble}
\title{Millimeter Wave MIMO Channel Estimation Based on Adaptive Compressed Sensing}

\author{\IEEEauthorblockN{Shu Sun and Theodore S. Rappaport}\\
\IEEEauthorblockA{NYU WIRELESS and Tandon School of Engineering, New York University, Brooklyn, NY, USA 11201\\
E-mail: \{ss7152,tsr\}@nyu.edu }

\thanks{Sponsorship for this work was provided by the NYU WIRELESS Industrial Affiliates program and NSF research grants 1320472, 1302336, and 1555332.}
}
\maketitle

\begin{abstract}
Multiple-input multiple-output (MIMO) systems are well suited for millimeter-wave (mmWave) wireless communications where large antenna arrays can be integrated in small form factors due to tiny wavelengths, thereby providing high array gains while supporting spatial multiplexing, beamforming, or antenna diversity. It has been shown that mmWave channels exhibit sparsity due to the limited number of dominant propagation paths, thus compressed sensing techniques can be leveraged to conduct channel estimation at mmWave frequencies. This paper presents a novel approach of constructing beamforming dictionary matrices for sparse channel estimation using the continuous basis pursuit (CBP) concept, and proposes two novel low-complexity algorithms to exploit channel sparsity for adaptively estimating multipath channel parameters in mmWave channels. We verify the performance of the proposed CBP-based beamforming dictionary and the two algorithms using a simulator built upon a three-dimensional mmWave statistical spatial channel model, NYUSIM, that is based on real-world propagation measurements. Simulation results show that the CBP-based dictionary offers substantially higher estimation accuracy and greater spectral efficiency than the grid-based counterpart introduced by previous researchers, and the algorithms proposed here render better performance but require less computational effort compared with existing algorithms. 

\end{abstract}

\IEEEpeerreviewmaketitle
\section{Introduction}
Multiple-input multiple-output (MIMO) is a key enabling technology for current and future wireless communication systems\cite{And14,Adh14,Lib99,Ertel,Durgin98}, as it offers great spectral efficiency and robustness due to many spatial degrees of freedom. Millimeter wave (mmWave) frequencies have been envisioned as a promising candidate for the fifth-generation (5G) wireless communications\cite{Rap13:Access}. Two prominent advantages of the mmWave spectrum are the massive bandwidth available and the tiny wavelengths compared to conventional UHF (Ultra-High Frequency)/microwave bands, thus enabling dozens or even hundreds of antenna elements to be implemented at communication link ends within a reasonable physical form factor. This suggests that MIMO and mmWave technologies should be combined to provide higher data rates, higher spectrum efficiency, thus resulting in lower latency\cite{Sun14:CM}. 

Channel state information (CSI) is needed to design precoding and combining procedures at transmitters and receivers, and it can be obtained through channel estimation. Conventional MIMO channel estimation methods may not be applicable in mmWave systems because of the substantially greater number of antennas, hence new channel estimation methods are required\cite{Heath16}. Due to the sparsity feature of mmWave channels observed in~\cite{Rap13:Access,Samimi16_TMTT}, compressed sensing (CS) techniques\cite{Can05} can be leveraged to effectively estimate mmWave channels\cite{Ber14,Alk15}. Adaptive CS, as a branch of CS, yields better performance at low signal-to-noise ratios (SNRs) compared to standard CS techniques, and low SNRs are typical for mmWave systems before implementing beamforming gain\cite{Alk14}. Adaptive CS algorithms for mmWave antenna arrays were derived in\cite{Alk14} to estimate channel parameters for both single-path and multipath scenarios, and it was shown that the proposed channel estimation approaches could achieve comparable precoding gains compared with exhaustive training algorithms. Additionally, Destino \textit{et al.} proposed an adaptive-least absolute shrinkage and selection operator (A-LASSO) algorithm to estimate sparse massive MIMO channels~\cite{Destino15}. In~\cite{Malla16}, reweighted $l_1$ minimization was employed to realize sparsity enhancement based on basis pursuit denoising. The authors of~\cite{Gao16} demonstrated a CS-based channel estimation algorithm for mmWave massive MIMO channels in ultra-dense networks, in conjunction with non-orthogonal pilots transmitted by  small-cell base stations (BSs).

In this paper, we propose an enhanced approach of creating the beamforming dictionary matrices for mmWave MIMO channel estimation in comparison with the one introduced in\cite{Alk14}, based on adaptive CS concepts. The main novelty of the proposed method here is the adoption of the continuous basis pursuit (CBP) method instead of the conventional grid-based approach to build beamforming dictionary matrices. This paper shows that the proposed dictionary can significantly improve the estimation accuracy, i.e., reduce the probability of estimation error, of angles of departure (AoDs) and angles of arrival (AoAs). Furthermore, built on the CBP-based dictionary, two new multipath channel estimation algorithms are proposed that have lower computational complexity compared to the one introduced in\cite{Alk14}, while offering better estimation accuracy for various signal sparsities. A three-dimensional (3D) statistical spatial channel model (SSCM)-based simulator, NYUSIM~\cite{Samimi16_TMTT,Sun17_NYUSIM}, developed for mmWave systems from extensive real-world propagation measurements, was used in the simulation to investigate the performance of the proposed algorithms.

The following notations are used throughout this paper. The boldface capital letter $\bold{X}$ and the boldface small letter $\bold{x}$ denote a matrix and a vector, respectively; A small italic letter $x$ denotes a scalar; $\mathbb{C}$ represents the set of complex numbers; $\mathbb{N}$ denotes the set of natural numbers; $||\bold{X}||_{F}$ denotes the Frobenius norm of $\bold{X}$; The conjugate, transpose, Hermitian, and Moore-Penrose pseudo inverse of $\bold{X}$ are represented by $\bold{X}^*$, $\bold{X}^T$, $\bold{X}^H$, and $\bold{X}^{\dagger}$, respectively; $\Tr(\bold{X})$ and $\vec(\bold{X})$ indicate the trace and vectorization of $\bold{X}$, respectively; The Hadamard, Kronecker and Khatri-Rao products between two matrices are denoted by $\circ$, $\otimes$, and $\ast$, respectively. 

\section{System Model}
Let us consider a BS equipped with $N_{\BS}$ antennas and $N_{\RF}$ RF chains communicating with a mobile station (MS) with $N_{\MS}$ antennas and $N_{\RF}$ RF chains, where $N_{\RF}\leq N_{\MS}\leq N_{\BS}$. We intentionally do not consider system interference issues, such as co-channel interference from other BSs and MSs, because of the limited interference found in directional mmWave channels~\cite{Heath16}, and also, the focus of this paper is to quantify and compare the performance of channel estimation methods in a single link. System aspects are ongoing research topics. In the channel estimation stage, the BS employs $M_{BS}$ beamforming vectors to transmit $M_{BS}$ symbols, while the MS utilizes $M_{MS}$ combining vectors to combine the received signal. The BS is assumed to implement analog/digital hybrid precoding with a precoding matrix $\bold{F}=\bold{F}_{\RF}\bold{F}_{\BB}$, where $\bold{F}_{\RF}\in\mathbb{C}^{N_{\BS}\times N_{\RF}}$ and $\bold{F}_{\BB}\in\mathbb{C}^{N_{\RF}\times M_{\BS}}$ denote the RF and baseband precoding matrices, respectively. Similarly, at the MS, the combiner $\bold{W}$ also consists of RF and baseband combiners represented by $\bold{W}_{\RF}\in\mathbb{C}^{N_{\MS}\times N_{\RF}}$ and $\bold{W}_{\BB}\in\mathbb{C}^{N_{\RF}\times M_{\MS}}$, respectively. The received signal at the MS is given by
\begin{equation}\label{Y1}
\bold{Y}=\bold{W}^H\bold{H}\bold{F}\bold{S}+\bold{Q}
\end{equation}

\noindent where $\bold{H}\in\mathbb{C}^{N_{\MS}\times N_{\BS}}$ denotes the channel matrix, $\bold{S}\in\mathbb{C}^{M_{\BS}\times M_{\BS}}$ is a diagonal matrix containing the $M_{\BS}$ transmitted symbols, and $\bold{Q}\in\mathbb{C}^{M_{\MS}\times M_{\BS}}$ represents the complex Gaussian noise. The design of analog/digital hybrid precoding and combining matrices have been extensively investigated\cite{Aya14,Yu16}, and we defer this topic to future work and focus on channel estimation in this paper. Additionally, although CSI can also be obtained by uplink training and channel reciprocity in time-division duplexing (TDD) systems, we focus on the downlink training in this paper since channel reciprocity usually does not hold for frequency-division duplexing (FDD) systems, and even in TDD systems if there exist non-linear devices that are not self-calibrated so as to incur non-reciprocal effects. 

The mmWave channel can be approximated by a geometric channel model with $L$ scatterers due to its limited scattering feature\cite{Rap13:Access,RapGut13,Rap15}, and the channel matrix can be written as
\begin{equation}\label{H}
\bold{H}=\sqrt{\frac{N_{\BS}N_{\MS}}{L}}\sum_{l=1}^{L}\alpha_l\bold{a}_{\MS}(\varphi_l,\vartheta_l)\bold{a}_{\BS}^H(\phi_l,\theta_l)
\end{equation}

\noindent where $\alpha_l$ is the complex gain of the $l^{th}$ path between the BS and MS  including the path loss, where a path refers to a cluster of multipath components traveling closely in time and/or spatial domains, $\phi_l, \varphi_l\in [0, 2\pi)$ are the azimuth AoD and AoA of the $l^{th}$ path, $\theta_l, \vartheta_l\in [-\pi/2, \pi/2]$ are the elevation AoD and AoA. $\bold{a}_{\BS}(\phi_l,\theta_l)$ and $\bold{a}_{\MS}(\varphi_l,\vartheta_l)$ are the antenna array response vectors at the BS and MS, respectively. The NYUSIM simulator produces a wide range of sample ensembles for~\eqref{H} and incorporates multiple antenna elements and physical arrays including uniform linear arrays (ULAs)~\cite{Sun17_NYUSIM}. Using a ULA, the array response vector can be expressed as (take the BS for example)
\begin{equation}
\bold{a}_{\BS}(\phi_l)=\frac{1}{\sqrt{N_{\BS}}}[1,e^{j\frac{2\pi}{\lambda}dcos(\phi_l)},\cdots, e^{j(N_{\BS}-1)\frac{2\pi}{\lambda}dcos(\phi_l)}]
\end{equation}

\noindent where the incident angle is defined as $0$ if the beam is parallel with the array direction, $\lambda$ denotes the carrier wavelength, and $d$ is the spacing between adjacent antenna elements. 

\section{Formulation of the mmWave Channel Estimation Problem}\label{formulation}
Considering the mmWave channel matrix given by~\eqref{H}, estimating the channel is equivalent to estimating the AoD, AoA, and path gain of each path, and training precoders and combiners are necessary to conduct the channel estimation. The mmWave channel estimation can be formulated as a sparse problem due to its limited dominant paths, e.g., on average 1 to 6 time clusters and 2 to 3 spatial lobes were found from real-world measurements using a 10 dB down threshold, as presented in~\cite{Samimi16_TMTT}. Therefore, some insights can be extracted from the CS theory. Assuming all transmitted symbols are equal for the estimation phase, i.e., $\bold{S}=\sqrt{P}\bold{I}_{M_{\BS}}$ ($P$ is the average power per transmission) and by vectorizing the received signal $\bold{Y}$ in~\eqref{Y1} to $\bold{y}$, we can approximate the received signal with a sparse formulation as follows\cite{Alk14}
\begin{equation}\label{y}
\begin{split}
\textbf{y}&=\sqrt{P}\vec(\bold{W}^H\bold{H}\bold{F})+\vec(\bold{Q})\\
&=\sqrt{P}(\textbf{F}^T\otimes\textbf{W}^H)\vec(\bold{H})+\textbf{n}_\Q\\
&=\sqrt{P}(\textbf{F}^T\otimes\textbf{W}^H)(\bold{A}_{\BS,\D}^*\ast\bold{A}_{\MS,\D})\textbf{z}+\textbf{n}_\Q\\
&=\sqrt{P}(\textbf{F}^T\textbf{A}_{\BS,\D}^*\otimes\textbf{W}^H\textbf{A}_{\MS,\D})\textbf{z}+\textbf{n}_\Q\\
&=\sqrt{P}\textbf{F}^T\textbf{A}_{\BS,\D}^*\textbf{z}_{\BS}\otimes\textbf{W}^H\textbf{A}_{\MS,\D}\textbf{z}_{\MS}+\textbf{n}_\Q
\end{split}
\end{equation}

\noindent where $\bold{A}_{\BS,\D}$ and $\bold{A}_{\MS,\D}$ denote the beamforming dictionary matrices at the BS and MS, respectively. $\textbf{z}_{\BS}\in\mathbb{C}^{N\times 1}$ and $\textbf{z}_{\MS}\in\mathbb{C}^{N\times 1}$ are two sparse vectors that have non-zero elements in the locations associated with the dominant paths, with $N$ denoting the number of measurements in the channel estimation stage, and $\bold{z}=\bold{z}_{\BS}\ast\bold{z}_{\MS}$. 

A beamforming dictionary based on angle quantization was proposed in\cite{Alk14}, where the AoDs and AoAs were assumed to be taken from a uniform grid of $N$ points with $N\gg L$ where $L$ denotes the number of paths, and the resulting dictionary matrix is expressed as (take the BS side for example, the MS dictionary matrix can be derived similarly)
\begin{equation}
\bold{A}_{\BS,\D}=[\bold{a}_{\BS}(\bar{\phi}_1),\cdots, \textbf{a}_{\BS}(\bar{\phi}_N)]
\end{equation}

\noindent where $\bold{a}_{\BS}(\bar{\phi}_n)$ ($n=1, ..., N$) denotes the BS array response vector for the grid point $\bar{\phi}_n$. 

Given that the true continuous-domain AoDs and AoAs may lie off the center of the grid bins, the grid representation in this case will destroy the sparsity of the signal and result in the so-called “basis mismatch”~\cite{Eka11}. This can be mitigated to a certain extent by finer discretization of the grid, but that may lead to higher computation time and higher mutual coherence of the sensing matrix, thus becoming less effective for sparse signal recovery\cite{Can05}. There are several approaches to mitigate the basis mismatch problem. One promising approach, named continuous basis pursuit (CBP), is proposed in\cite{Eka11}, where one type of CBP is implemented with first-order Taylor interpolator, which will be demonstrated shortly. 

Since the antenna array factor $\textbf{a}(\phi)$ is a continuous and smooth function of $\phi$, it can be approximated by linearly combining $\textbf{a}(\phi_k)$ and the derivative of $\textbf{a}(\phi)$ at the point $\phi_k$ via a first-order Taylor expansion:
\begin{equation}\label{Taylor}
\textbf{a}(\phi)=\textbf{a}(\phi_k)+(\phi-\phi_k)\frac{\partial \textbf{a}(\phi)}{\partial \phi}\biggr\rvert_{\phi_k}+\mathcal{O}((\phi-\phi_k)^2)
\end{equation}

\noindent where $\phi_k = 2\pi(k-1)/N$ is the grid-point with minimal distance from $\phi$. This motivates a dictionary consisting of the original discretized array factors $\textbf{a}(\phi)$ and its derivatives $\frac{\partial \textbf{a}(\phi)}{\partial \phi}$, i.e., $\textbf{a}(\phi)$ and $\frac{\partial \textbf{a}(\phi)}{\partial \phi}$ can be regarded as two sets of basis for the dictionary. Therefore, the entire basis for the proposed dictionary matrix can be formulated as
\begin{equation}
\bold{B}_{\BS}=[\textbf{a}_{\BS}(\phi_1),\cdots, \textbf{a}_{\BS}(\phi_N),\textbf{b}_{\BS}(\phi_1),\cdots, \textbf{b}_{\BS}(\phi_N)]
\end{equation}

\noindent where $\textbf{b}_{\BS}(\phi_n)=\frac{\partial \textbf{a}_{\BS}(\phi)}{\partial \phi}\bigr\rvert_{\phi_n}$, and the corresponding interpolator is given by
\begin{equation}
\bold{t}_{\BS} = [\underbrace{1,\cdots, 1}_{N}, \underbrace{\Delta\phi,\cdots, \Delta\phi}_{N}]^T
\end{equation}

\noindent where $\Delta\phi$ denotes the angle offset from the angles on the grid, and $|\Delta\phi|\leq\frac{\pi}{N}$. The proposed dictionary is hence written as
\begin{equation}\label{A_D_new}
\begin{split}
\tilde{\bold{A}}_{\BS,\D}=\bold{B}_{\BS}\bold{t}_{\BS}=&[\textbf{a}_{\BS}(\phi_1),\cdots, \textbf{a}_{\BS}(\phi_N),\\
&\Delta\phi\textbf{b}_{\BS}(\phi_1),\cdots, \Delta\phi\textbf{b}_{\BS}(\phi_N)]
\end{split}
\end{equation}


\section{Multi-Resolution Hierarchical Codebook}
The proposed hierarchical beamforming codebook is composed of $S$ levels, where each level contains beamforming vectors with a certain beamwidth that covers certain angular regions. Due to the symmetry of the antenna pattern of a ULA, if a beam covers an azimuth angle range of $[\phi_a, \phi_b]$, then it also covers $2\pi-[\phi_a, \phi_b]$. In each codebook level $s$, the beamforming vectors are divided into $K^{s-1}$ subsets, each of which contains $K$ beamforming vectors. Each of these $K$ beamforming vectors is designed such that it has an almost equal projection on the vectors $\bold{a}_{\BS}(\bar{\phi})$, where $\bar{\phi}$ denotes the angle range covered by this beamforming vector, and zero projection on the array response vectors corresponding to other angles. Note that there is no strict constraint on the number of sectors $K$ at each stage, yet considering practical angle-searching time, $K=$~3 or 4 is a reasonable choice. Once the value of $K$ is defined, the total number of estimation measurements $N$ is $2K^S$. The value of $N$ should be minimized while guaranteeing the successful estimation of angles, thus $S$ should be neither too large nor too small. Through simulations, it is found that $S=3,K=4~(N=128)$ and $S=4,K=3~(N=162)$ are two sensible combinations.


In each codebook level $s$ and subset $k$, the $m^\text{th}$ column of the beamforming vector $[\bold{F}_{(s,k)}]_{:,m}, m=1,..., K$ in the codebook $\mathcal{F}$ is designed such that

$[\bold{F}_{(s,k)}]_{:,m}^H\bold{a}_{\BS}(\bar{\phi}_u)=\left\{ \begin{array}{rcl}C, & \for~\bar{\phi}_u\in \mathcal{\Phi}_{s,k,m}^{\BS} \\ 0, & \otherwise
\end{array}\right.$

\begin{equation}\label{Fb}
[\bold{F}_{(s,k)}]_{:,m}^H\bold{b}_{\BS}(\bar{\phi}_u)=0, \forall~\bar{\phi}_u
\end{equation}

\noindent with 
\begin{equation}\label{Phi}
\begin{split}
\mathcal{\Phi}_{s,k,m}^{\BS}=&\Big[\frac{\pi}{K^s}\big(K(k_s^{\BS}-1)+m_{\BS}-1\big), \frac{\pi}{K^s}(K(k_s^{\BS}-1)\\
&+m_{\BS})\Big]\cup \Big[2\pi-\frac{\pi}{K^s}(K(k_s^{\BS}-1)+m_{\BS}), \\
&2\pi-\frac{\pi}{K^s}\big(K(k_s^{\BS}-1)+m_{\BS}-1\big)\Big]
\end{split}
\end{equation}

\noindent where $C$ is a constant such that each $\bold{F}_{(s,k)}$ has a Frobenius norm of $K$.  The fact that the product of $[\bold{F}_{(s,k)}]_{:,m}^H$ and $\bold{b}_{\BS}(\bar{\phi}_u)$ is zero in~\eqref{Fb} can be derived from~\eqref{Taylor} to~\eqref{A_D_new}. The matrix $\bold{F}_{(s,k)}$ hence equals the product of the pseudo-inverse of $\tilde{\bold{A}}_{\BS,\D}$ and the $(k\times K-(K-1))^\text{th}$ to $(k\times K)^\text{th}$ columns of the angle coverage matrix $\bold{G}_{(s)}$ with its $m^\text{th}$ column given by~\eqref{G}: 
\begin{equation}\label{G}
[\bold{G}_{(s)}]_{:,m} = [\underbrace{C\rq{}s~\and~0\rq{}s}_{N}, \underbrace{0,\cdots, 0}_{N}]^T
\end{equation}

\noindent where $C$\rq{}s are in the locations $\mathcal{\Phi}_{s,k,m}^{\BS}$. The combining matrix $\bold{W}_{(s,k)}$ in the codebook $\mathcal{W}$ at the receiver can be designed in a similar manner. It is noteworthy that the difference between the angle coverage matrix $\bold{G}_{(s)}$ in~\cite{Alk14} and the one proposed here is that the $m^\text{th}$ column of the former contains only the first $N$ rows without the last $N$ 0\rq{}s in~\eqref{G}, i.e., the former did not force $[\bold{F}_{(s,k)}]_{:,m}^H\bold{b}_{\BS}(\bar{\phi}_u)$ to be zero, hence failing to alleviate the leakage incurred by angle quantization. 

\begin{figure}
\centering
 \includegraphics[width=3.4in]{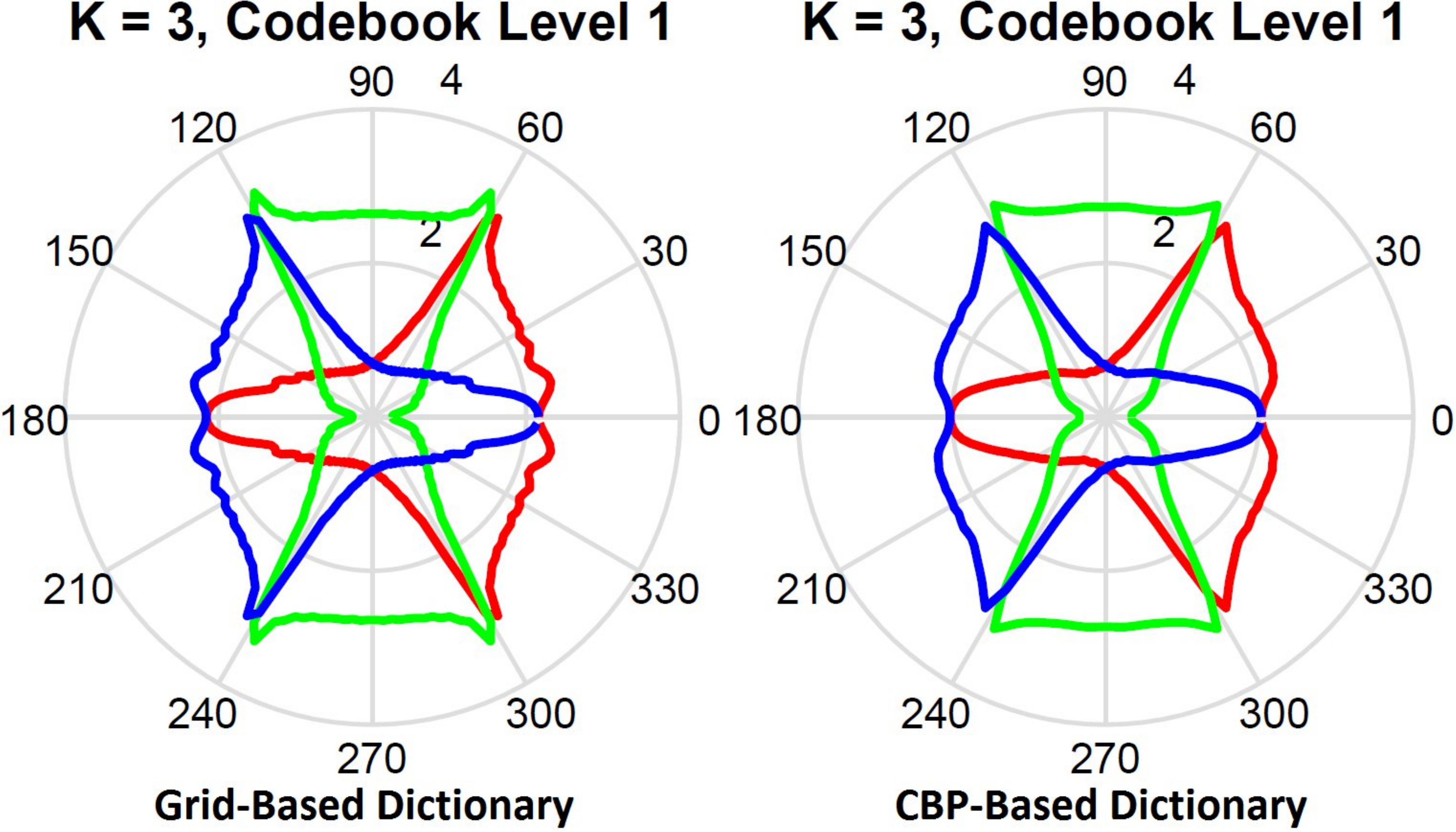}
    \caption{Beam patterns of the beamforming vectors in the first codebook level of an example hierarchical codebook using the grid-based and CBP-based dictionaries with $N$ = 162, $K$ = 3.}
    \label{fig:CL_K3}
\end{figure}

Fig.~\ref{fig:CL_K3} illustrates the beam patterns of the beamforming vectors in the first codebook level of an example hierarchical codebook introduced in\cite{Alk14} and the hierarchical codebook proposed in this paper with $N$ = 162 and $K$ = 3. Comparing the two beam patterns, we can see that the codebook generated using the CBP-based dictionary $\tilde{\bold{A}}_{\BS,\D}$ in~\eqref{A_D_new} produces a smoother pattern contour in contrast to that yielded by the codebook introduced in\cite{Alk14}, namely, the beams associated with $\tilde{\bold{A}}_{\BS,\D}$ are able to cover the intended angle ranges more evenly. Due to the more uniform projection on the targeted angle region, the beamforming vectors generated using $\tilde{\bold{A}}_{\BS,\D}$ can mitigate the leakage induced by angle quantization, thus improving the angle estimation accuracy, as will be shown later. 

\section{Adaptive estimation algorithms for mmWave MIMO channels}
For single-path channels, there is only one non-zero element in the vector $\bold{z}$ in~\eqref{y}. To effectively estimate the location of this non-zero element, and consequently the corresponding AoD, AoA, and path gain, the following algorithm, which is an improved version of Algorithm 1 in\cite{Alk14}, is used in conjunction with the innovative CBP-based dictionary matrices. 

\begin{algorithm}
\caption{Adaptive Estimation Algorithm for Single-Path mmWave MIMO channels}
\label{Al1}
\begin{algorithmic}[1]
\Require{$K$, $S$, codebooks $\mathcal{F}$ and $\mathcal{W}$, $N=2K^S$}
\State \textbf{Initialization}: $k_{1}^{\BS}=1$, $k_{1}^{\MS}=1$
\For{$s\leq S$}
\For{$m_{\BS}\leq K$}
\State BS uses $[\textbf{F}_{(s,k_s^{\BS})}]_{:,m_{\BS}}$
\For{$m_{\MS}\leq K$}
\State MS uses $[\textbf{W}_{(s,k_s^{\MS})}]_{:,m_{\MS}}$
\EndFor~{$m_{\MS}\leq K$}
\EndFor~{$m_{\BS}\leq K$}
\EndFor~{$s\leq S$}
\For{$s\leq S$}
\State $\textbf{Y}_{(s)}=\sqrt{P_s}[\textbf{W}_{(s,k_s^{\MS})}]^H\textbf{H}[\textbf{F}_{(s,k_s^{\BS})}]+\textbf{Q}$
\State $\{m_{\BS}^*,m_{\MS}^*\}=\underset{\forall m_{\BS},m_{\MS}=1,...,K}{\operatorname{argmax}}[\textbf{Y}_{(s)}\circ \textbf{Y}_{(s)}^*]_{m_{\MS},m_{BS}}$
\State $\hat{\phi}_{\can}\in \mathcal{\Phi}_{s,k,m}^{\BS}$, $\hat{\varphi}_{\can}\in \mathcal{\Phi}_{s,k,m}^{\MS}$ \% $\mathcal{\Phi}_{s,k,m}^{\BS}$ is given by Eq.~\eqref{Phi}, and $\mathcal{\Phi}_{s,k,m}^{\MS}$ can be calculated similarly
\State $k_{s+1}^{\BS}=K(k_s^{\BS}-1)+m_{\BS}^*$, $k_{s+1}^{\MS}=K(k_s^{\MS}-1)+m_{\MS}^*,$~
\EndFor~{$s\leq S$}
\State $\bold{A}_{\can,\BS}=[\bold{a}_{\BS}(\hat{\phi}_{\can})]$ \% Antenna array matrix for the candidate AoDs

\noindent $\bold{A}_{\can,\MS}=[\bold{a}_{\MS}(\hat{\varphi}_{\can})]$ \% Antenna array matrix for the candidate AoAs

\noindent $\bold{Z}=\bold{A}_{\can,\MS}^H\bold{H}\bold{A}_{\can,\BS}+\bold{Q}$ \% Received signal matrix corresponding to the candidate AoDs and AoAs

\noindent $(\hat{\phi}, \hat{\varphi})=\argmax~\bold{Z}\circ \bold{Z}^*$ \% Finding the optimal AoD and AoA that maximize the Hadamard product of the received signal matrix

\noindent $\hat{\alpha}=\sqrt{\bold{Z}_{(\hat{\phi}, \hat{\varphi})}\circ \bold{Z}^*_{(\hat{\phi}, \hat{\varphi})}}/(N_{\BS}*N_{\MS})$ \% Estimated path gain magnitude associated with the estimated AoD and AoA
\Ensure{$\hat{\phi}, \hat{\varphi}, \hat{\alpha}$}
\end{algorithmic}
\end{algorithm}

Algorithm 1 operates as follows. In the initial stage, the BS uses the training precoding vectors of the first level of the codebook $\mathcal{F}$. For each of those vectors, the MS uses the measurement vectors of the first level of $\mathcal{W}$ to combine the received signal. After the precoding-measurement steps of this stage, the MS compares the power of the received signals to determine the one with the maximum received power. As each one of the precoding/measurement vectors is associated with a certain range of the quantized AoA/AoD, the operation of the first stage divides the entire angle range $[0,~2\pi)$ into $K$ partitions, and compares the power of the sum of each of them. Hence, the selection of the strongest received signal implies the selection of the range of the quantized AoA/AoD that is highly likely to contain the single path of the channel. The output of the maximum power is then used to determine the subsets of the beamforming vectors of level $s+1~(1\leq s\leq S-1)$ of $\mathcal{F}$ and $\mathcal{W}$ to be used in the next stage. Since $N$ must be even multiples of $K$ in order to construct the precoding and measurement codebooks, there are two possible ranges of AoD/AoA selected out after Step 15 of Algorithm 1, which are denoted as $\hat{\phi}_{\can}$ and $\hat{\varphi}_{\can}$. Step 16 is aimed at ``filtering'' out the AoD/AoA from these two ranges. The MS then feeds back the selected subset of the BS precoders to the BS to use it in the next stage, which needs only $\log_2 K$ bits.


Based on Algorithm 1, two low-complexity algorithms for estimating multipath channels are established, as explained below.
\begin{algorithm}
\caption{Adaptive Estimation Algorithm for Multipath mmWave MIMO channels}
\label{Al2}
\begin{algorithmic}[1]
\Require{$K$, $S$, codebooks $\mathcal{F}$ and $\mathcal{W}$, $N=2K^S$}
\State \textbf{Initialization}: $\mathcal{I}_{(:,1)}^{\BS}=[1,...,1]^T,~\mathcal{I}_{(:,1)}^{\MS}=[1,...,1]^T$, where $\mathcal{I}^{\BS}\in\mathbb{N}^{L\times S},~\mathcal{I}^{\MS}\in\mathbb{N}^{L\times S}$ 
\State Use Algorithm 1 to detect the AoD, AoA, and path gain for the first strongest path 
\State Repeat Algorithm 1 for the $l^{th}~(2\leq l\leq L)$ path until Step 11 in Algorithm 1
\State For the $s^{th}$ stage in the $i^{th}~(2\leq i\leq L)$ iteration, project out previous path contributions one path by one path

\noindent $\textbf{Y}_{(s)}=\sqrt{P_s}[\textbf{W}_{(s,k_s^{\MS})}]^H\textbf{H}[\textbf{F}_{(s,k_s^{\BS})}]+\textbf{Q}$

\noindent $\bold{y}_{(s)}=\vec(\bold{Y}_{(s)})$

\noindent $\bold{V}_{(i,s)}=\bold{F}^T_{(s,\mathcal{I}_{(i,s)}^{\BS})}[\tilde{\bold{A}}_{\BS,\D}]^*_{:,\mathcal{I}_{(i,s)}^{\BS}}\otimes\bold{W}^H_{(s,\mathcal{I}_{(i,s)}^{\MS})}[\tilde{\bold{A}}_{\MS,\D}]_{:,\mathcal{I}_{(i,s)}^{\MS}}$ \% Calculating the contribution of previous paths in the form of Eq.~\eqref{y}

\noindent $\bold{y}_{(s)}=\bold{y}_{(s)}-\bold{V}_{(i,s)}\bold{V}_{(i,s)}^{\dagger}\bold{y}_{(s)}$

\State Convert $\bold{y}_{(s)}$ to the matrix form $\bold{Y}_{(s)}$
\State Repeat Algorithm 1 from Step 12 to obtain the AoD, AoA, and path gain for the $i^{th}$ strongest path until all the $L$ paths are estimated
\Ensure{AoDs, AoAs, and path gains for the $L$ dominant paths}
\end{algorithmic}
\end{algorithm}

In Algorithm 2, $\mathcal{I}_{(i,s)}^{\BS}$ and $\mathcal{I}_{(i,s)}^{\MS}$ contain the precoding and measurement matrix indexes of the $i^{th}$ path in the $s^{th}$ stage, respectively. Algorithm 2 operates as follows: A procedure similar to Algorithm 1 is utilized to detect the first strongest path. The indexes of the beamforming matrices corresponding to the previous detected $l~(1\leq l\leq L-1)$ paths are stored and used in later iterations.  Note that in each stage $s$ from the second iteration on, the contribution of the paths that have already been estimated in previous iterations are projected out one path by one path before determining the new promising AoD/AoA ranges. In the next stage $s+1$, two AoD/AoA ranges are selected for further refinement, i.e., the one selected at stage $s$ of this iteration, and the one selected by the preceding path at stage $s+1$ of the previous iteration. The algorithm makes $L$ outer iterations to estimate $L$ paths. Thanks to the sparse nature of mmWave channels, the number of dominant paths is usually limited, which means the total number of precoding-measurement steps will not be dramatically larger compared to the single-path case. 

\begin{algorithm}
\caption{Adaptive Estimation Algorithm for Multipath mmWave MIMO channels}
\label{Al3}
\begin{algorithmic}[1]
\Require{$K$, $S$, codebooks $\mathcal{F}$ and $\mathcal{W}$, $N=2K^S$}
\State \textbf{Initialization}: $\mathcal{I}_{(:,1)}^{\BS}=[1,...,1]^T,~\mathcal{I}_{(:,1)}^{\MS}=[1,...,1]^T$, where $\mathcal{I}^{\BS}\in\mathbb{N}^{L\times S},~\mathcal{I}^{\MS}\in\mathbb{N}^{L\times S}$ 
\State Use Algorithm 1 to detect the AoD, AoA, and path gain for the first strongest path 
\State Repeat Algorithm 1 for the $l^{th}~(2\leq l\leq L)$ path until Step 11 in Algorithm 1
\State For the $s^{th}$ stage in the $i^{th}~(2\leq i\leq L)$ iteration, project out previous path contributions simultaneously

\noindent $\hat{\bold{A}}_{\BS}=[\bold{a}_{\BS}(\hat{\phi})],~\hat{\bold{A}}_{\MS}=[\bold{a}_{\MS}(\hat{\varphi})]$ \% $\hat{\phi}$ and $\hat{\varphi}$ are the AoDs and AoAs of all the previously detected paths, respectively

\noindent $\textbf{Y}_{(s)}=\sqrt{P_s}[\textbf{W}_{(s,k_s^{\MS})}]^H\textbf{H}[\textbf{F}_{(s,k_s^{\BS})}]+\textbf{Q}$

\noindent $\bold{y}_{(s)}=\vec(\bold{Y}_{(s)})$

\noindent $\bold{V}_{(i,s)}=[\textbf{W}_{(s,k_s^{\MS})}]^H\hat{\bold{A}}_{\MS}\hat{\bold{A}}_{\BS}^H[\textbf{F}_{(s,k_s^{\BS})}]$ \% Calculating the contribution of previous paths in the form of Eq.~\eqref{Y1}

\noindent $\bold{v}_{(i,s)}=\vec(\bold{V}_{(i,s)})$

\noindent $\bold{y}_{(s)}=\bold{y}_{(s)}-\bold{v}_{(i,s)}\bold{v}_{(i,s)}^{\dagger}\bold{y}_{(s)}$

\State Convert $\bold{y}_{(s)}$ to the matrix form $\bold{Y}_{(s)}$
\State Repeat Algorithm 1 from Step 12 to obtain the AoD, AoA, and path gain for the $i^{th}$ strongest path until all the $L$ paths are estimated
\Ensure{AoDs, AoAs, and path gains for the $L$ dominant paths}
\end{algorithmic}
\end{algorithm}

Algorithm 3 is similar to Algorithm 2, but with an even lower complexity. The major difference between Algorithm 3 and Algorithm 2 stems from the way of projecting out previous path contributions: Algorithm 3 does not require storing the beamforming matrix indexes, but instead, it utilizes the antenna array response vectors associated with the estimated AoDs/AoAs to subtract out the contributions of previously detected paths simultaneously. Therefore, compared with Algorithm 2, Algorithm 3 results in less computation and storage cost, and a higher estimation speed (i.e., lower latency). 

When compared with the multipath channel estimation presented in\cite{Alk14}, the most prominent advantages of both Algorithm 2 and Algorithm 3 are that they do not require the re-design of multi-resolution beamforming codebooks for each stage when the number of dominant paths vary, and only a single path is selected in each stage instead of $L$ paths in\cite{Alk14}, thus substantially reducing the calculation and memory overhead.

\section{Simulation Results}
In this section, the performance of the proposed CBP-based dictionary and Algorithms 1, 2, and 3 are evaluated in terms of average probability of estimation error of AoDs and AoAs, and spectral efficiency, via numerical Monte Carlo simulations. The channel matrix takes the form of~\eqref{H}, where the path powers, phases, AoDs, and AoAs are generated using the 5G open-source wideband simulator, NYUSIM, as demonstrated in\cite{Samimi16_TMTT,Sun17_NYUSIM}. The channel model in NYUSIM utilizes the time-cluster-spatial-lobe (TCSL) concept, where for an RF bandwidth of 800 MHz, the number of TCs varies from 1 to 6 in a uniform manner, the number of subpaths per TC is uniformly distributed between 1 and 30, and the number of SLs follows the Poisson distribution with an upper bound of 5. The NYUSIM channel model is applicable to arbitrary frequencies from 500 MHz to 100 GHz, RF bandwidths between 0 and 800 MHz, various scenarios (urban microcell (UMi), urban macrocell (UMa), and rural macrocell (RMa)~\cite{Mac17b}), and a vast range of antenna beamwidths~\cite{Samimi16_TMTT,Sun17_NYUSIM}. 

ULAs are assumed at both the BS and MS with 64 and 32 antenna elements, respectively. All simulation results are averaged over 10,000 random channel realizations, with a carrier frequency of 28 GHz and an RF bandwidth of 800 MHz. In calculating spectral efficiency, eigen-beamforming is assumed at both the transmitter (with equal power allocation) and receiver. Other beamforming techniques can also be employed, and we found the performance of the beamforming dictionaries and algorithms are similar.  

\begin{figure}
\centering
 \includegraphics[width=3.4in]{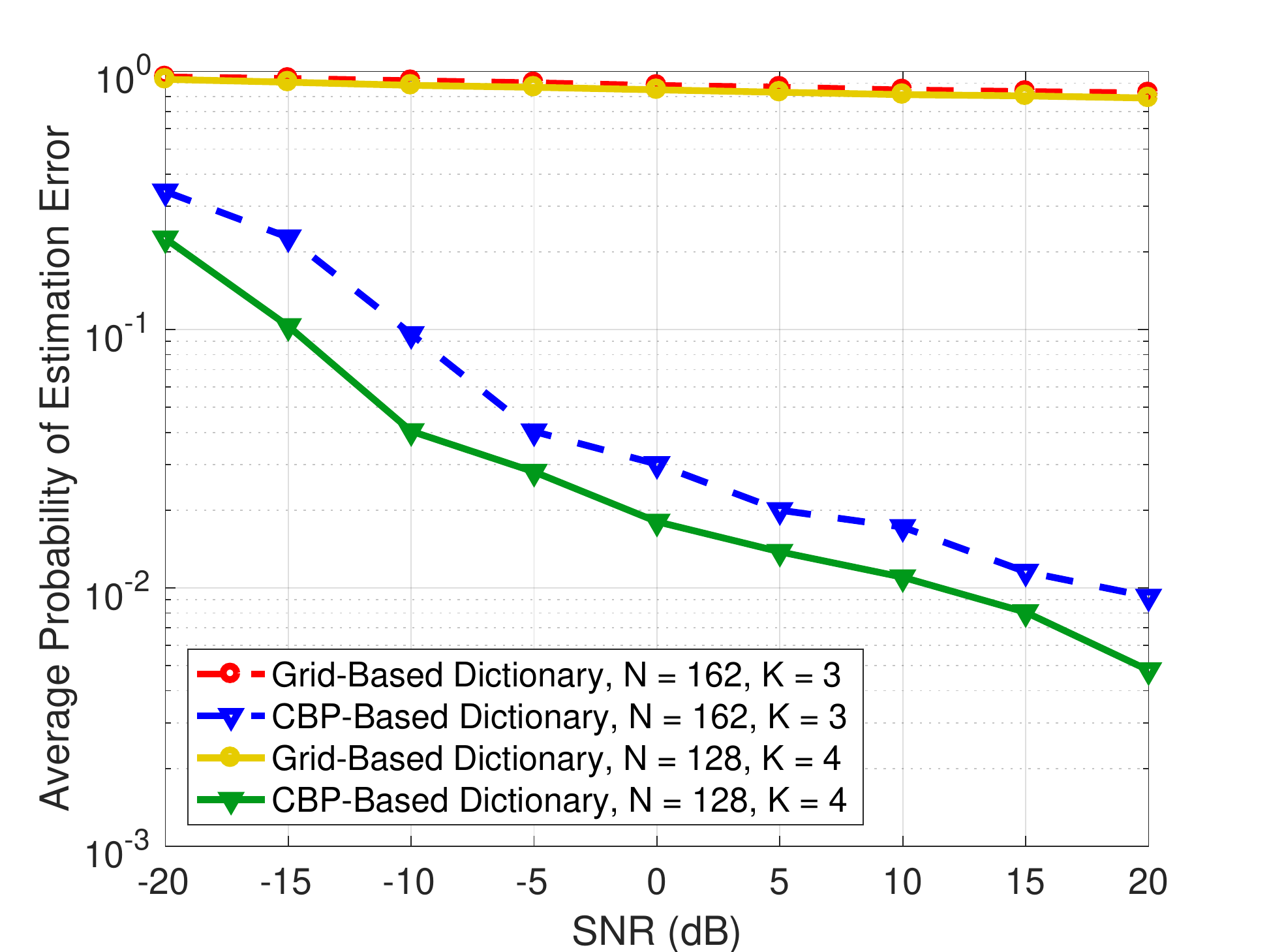}
    \caption{Average probability of error in estimating AoD/AoA for single-path channels, using both the grid-based dictionary and CBP-based dictionary.}
    \label{fig:Error_SNR}
\end{figure}

The simulated probabilities of estimation errors of AoDs and AoAs as a function of the average receive SNR, using Algorithm 1 and both grid-based and CBP-based dictionaries for single-path channels, are depicted in Fig.~\ref{fig:Error_SNR} for the cases of $N$ = 162, $K$ = 3, and $N$ = 128, $K$ = 4, which are found to yield the best performance via numerous trials. As shown by Fig.~\ref{fig:Error_SNR}, the CBP-based approach renders much smaller estimation errors, by up to two orders of magnitude. For the two cases considered in Fig.~\ref{fig:Error_SNR}, the grid-based method generates huge estimation error probability that is over 80\% even at an SNR of 20 dB; on the other hand, the estimation error probability of the CBP-based counterpart decreases rapidly with SNR, and is less than 0.5\% for $N$ = 128, $K$ = 4 and a 20 dB SNR. These results imply that the CBP-based approach is able to provide much better channel estimation accuracy with a small number of measurements compared to the conventional grid-based fashion, hence is worth using in mmWave MIMO systems for sparse channel estimation and signal recovery. 

\begin{figure}
\centering
 \includegraphics[width=3.4in]{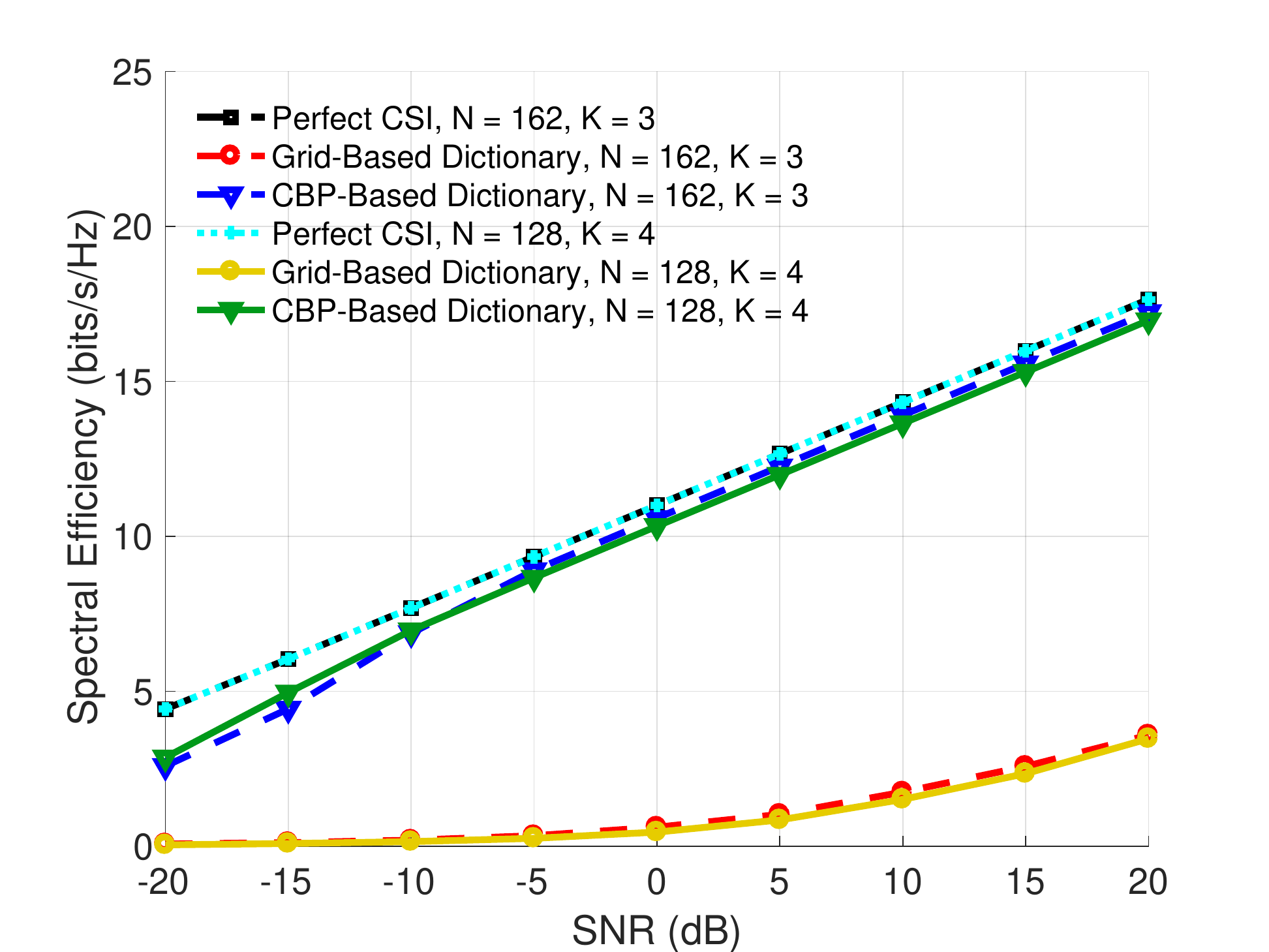}
    \caption{Average spectral efficiency for single-path channels for the cases of perfect CSI, grid-based dictionary and CBP-based dictionary.}
    \label{fig:SE_SNR_L1}
\end{figure}

To explicitly show the effect of estimation error on channel spectral efficiency using different beamforming dictionaries, we plot and compare the achievable spectral efficiency as a function of the average receive SNR for both the grid-based and CBP-based dictionaries for single-path channels, as well as the spectral efficiency with perfect CSI at the transmitter, for the case of $N$ = 162, $K$ = 3, and $N$ = 128, $K$ = 4, as described in Fig.~\ref{fig:SE_SNR_L1}. It is evident from Fig.~\ref{fig:SE_SNR_L1} that for both cases considered, the CBP-based dictionary yields much higher spectral efficiency, by about 2.7 bits/s/Hz to 13 bits/s/Hz, compared with the grid-based one over the entire SNR range of -20 dB to 20 dB. Furthermore, the CBP-based method achieves near-optimal performance over the SNRs spanning from 0 dB to 20 dB, with a gap of less than 0.7 bits/s/Hz.


\begin{figure}
\centering
 \includegraphics[width=3.4in]{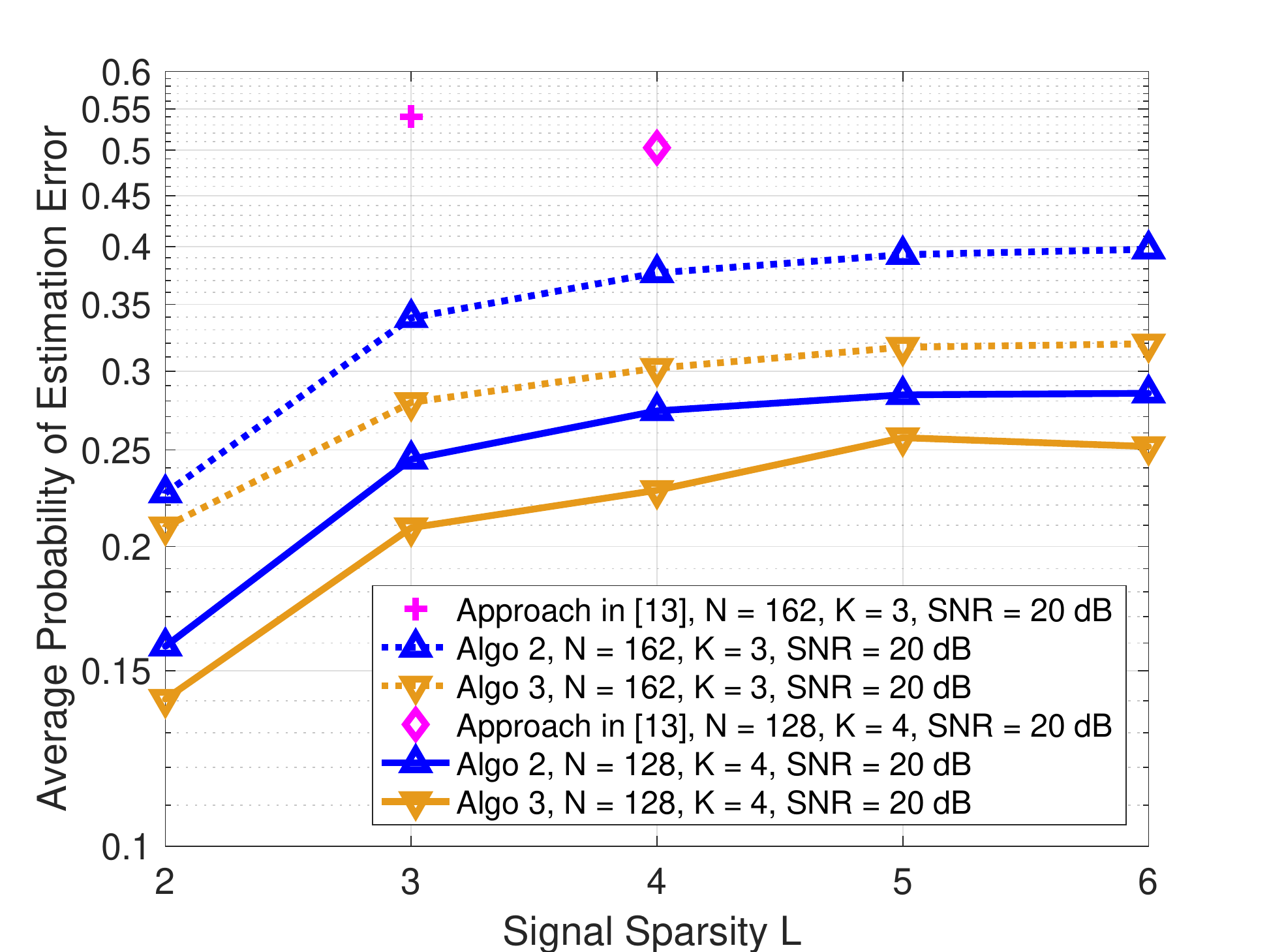}
    \caption{Average probability of error in estimating AoD/AoA for multipath channels using the CBP-based dictionary.}
    \label{fig:Error_L_K3K4_SNR20}
\end{figure}

Fig.~\ref{fig:Error_L_K3K4_SNR20} illustrates the average probability of error in estimating AoDs/AoAs for multipath channels with $N$ = 162, $K$ = 3, and $N$ = 128, $K$ = 4 for an average receive SNR of 20 dB, using proposed Algorithms 2  and 3 for two to six dominant paths, as well as Algorithm 2 in\cite{Alk14}. For the approach in\cite{Alk14}, since all $L$ paths have to be estimated simultaneously in a multipath channel, it does not work for $L<K$, thus no results are available for $L=2$ when $K=3$ or $4$. The SNR denotes the ratio of the total received power from \textit{all} paths to the noise power. As shown in Fig.~\ref{fig:Error_L_K3K4_SNR20}, both Algorithm 2 (Algo 2) and Algorithm 3 (Algo 3) produce lower estimation errors than the approach in\cite{Alk14} in both multipath-channel cases; for the case of $N=128$, $K=4$, Algorithm 3 yields the lowest estimation error, i.e., highest accuracy, and meanwhile enjoys the lowest computation expense among the three algorithms. In addition, the estimation error tends to increase more slowly and converge to a certain value as the number of dominant paths increase for all of the three algorithms. 


\begin{figure}
\centering
 \includegraphics[width=3.4in]{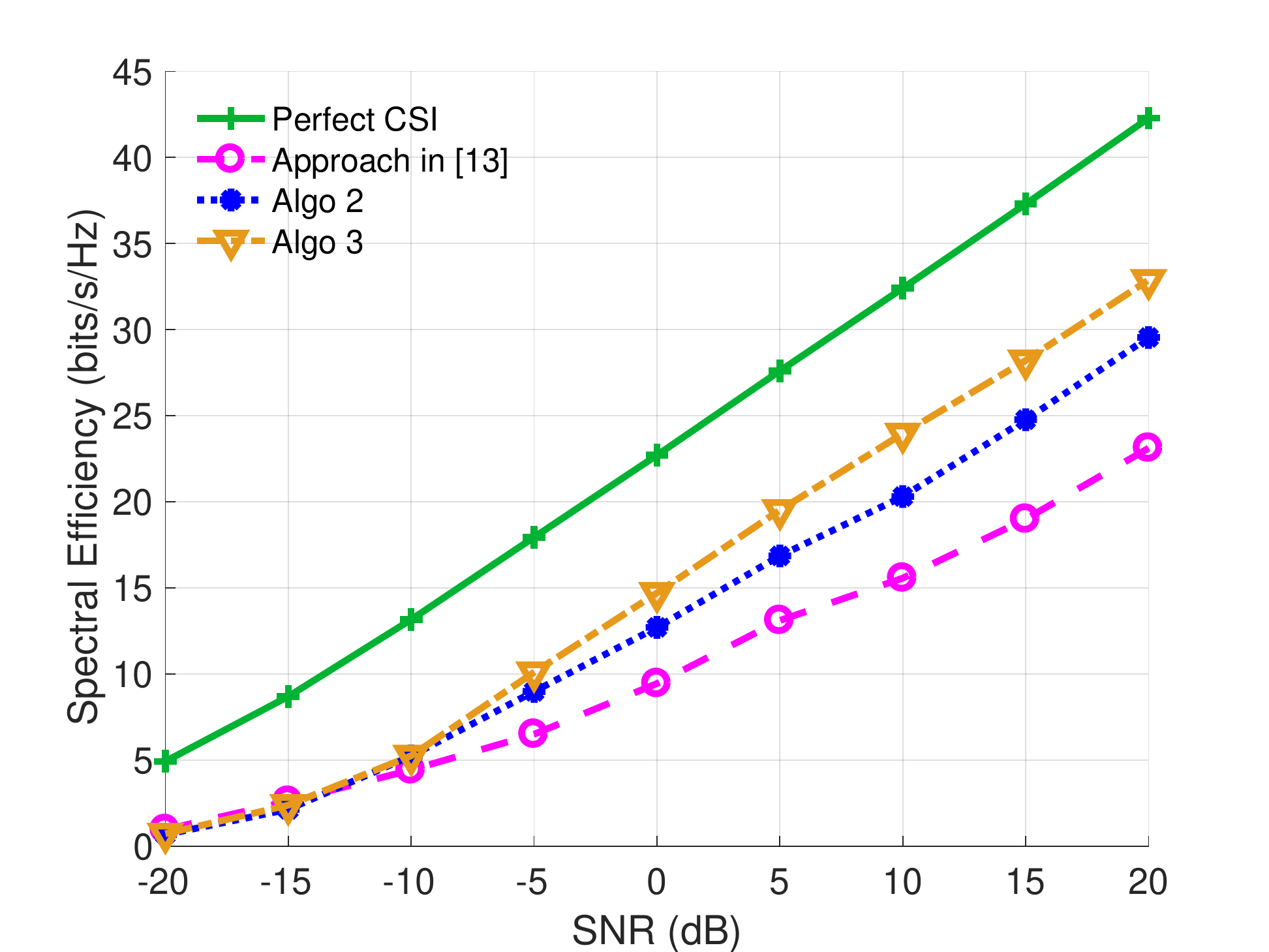}
    \caption{Average spectral efficiency for multipath channels with the CBP-based dictionary using the approach in~\cite{Alk14}, and Algorithms 2 and 3 proposed in this paper, with $N$ = 162, $K$ = 3, and $L$ = 3.}
    \label{fig:SE_SNR_K3_N162_L3}
\end{figure}

The spectral efficiency performance of the three algorithms above, with $N$ = 162, $K$ = 3, and $L$ = 3, is displayed in Fig.~\ref{fig:SE_SNR_K3_N162_L3}, which reveals the superiority of Algorithm 3 pertaining to spectral efficiency, followed by Algorithm 2, compared with the approach in\cite{Alk14}. For instance, at an SNR of 10 dB, Algorithms 2 and 3 yield around 5 and 8 more bits/s/Hz than the approach in\cite{Alk14}, respectively, and the discrepancies expand as the SNR ascends. The proposed algorithms work well for single-path channels, and significantly outperforms the approach in\cite{Alk14} method for multipath channels, although there is still a noticeable spectral efficiency gap compared to the perfect CSI case, due to the non-negligible angle estimation errors shown in Fig.~\ref{fig:Error_L_K3K4_SNR20}. Further work is needed to improve Algo 2 and Algo 3 to more effectively estimate multipath channels. 

\section{Conclusion}
Based on the concept of adaptive compressed sensing and by exploiting the sparsity of mmWave channels, in this paper, we presented an innovative approach for designing the precoding/measurement dictionary matrices, and two new low-complexity algorithms for estimating multipath channels. In contrast to the conventional grid-based method, the principle of CBP was leveraged in devising the beamforming dictionary matrices, which had lower mutual coherence due to the first-order Taylor interpolation, and was shown to be more beneficial for sparse signal reconstruction. 

Simulations were performed based on the open-source 5G channel simulator NYUSIM for broadband mmWave systems. Results show that the CBP-based dictionary renders up to over two orders of magnitude higher estimation accuracy (i.e., lower probability of estimation error) of AoDs and AoAs, and more than 12 bits/s/Hz higher spectral efficiency, with a small number of estimation measurements for single-path channels, as opposed to the grid-based approach, as shown in Figs.~\ref{fig:Error_SNR} and~\ref{fig:SE_SNR_L1}. Moreover, the newly proposed two algorithms, Algorithm 2 and Algorithm 3, can offer better estimation and spectral efficiency performance with lower computational complexity and time consumption for multipath channels, when compared with existing algorithms, as shown in Figs.~\ref{fig:Error_L_K3K4_SNR20} and~\ref{fig:SE_SNR_K3_N162_L3}.

Interesting extensions to this work will be to improve the multipath estimation algorithms to make them more effective, and to extend the multipath estimation algorithms to the case where the number of dominant paths is unknown, as well as to implement the proposed dictionary matrices and algorithms to other types of antenna arrays such as 2D arrays.


\ifCLASSOPTIONcaptionsoff
  \newpage
\fi

\bibliographystyle{IEEEtran}
\bibliography{bibliography}

\begin{thebibliography}{10}
\providecommand{\url}[1]{#1}
\csname url@samestyle\endcsname
\providecommand{\newblock}{\relax}
\providecommand{\bibinfo}[2]{#2}
\providecommand{\BIBentrySTDinterwordspacing}{\spaceskip=0pt\relax}
\providecommand{\BIBentryALTinterwordstretchfactor}{4}
\providecommand{\BIBentryALTinterwordspacing}{\spaceskip=\fontdimen2\font plus
\BIBentryALTinterwordstretchfactor\fontdimen3\font minus
  \fontdimen4\font\relax}
\providecommand{\BIBforeignlanguage}[2]{{%
\expandafter\ifx\csname l@#1\endcsname\relax
\typeout{** WARNING: IEEEtran.bst: No hyphenation pattern has been}%
\typeout{** loaded for the language `#1'. Using the pattern for}%
\typeout{** the default language instead.}%
\else
\language=\csname l@#1\endcsname
\fi
#2}}
\providecommand{\BIBdecl}{\relax}
\BIBdecl

\bibitem{And14}
\text{J. G. Andrews} \textit{et al.}, ``What will 5\text{G} be?'' \emph{IEEE
  Journal on Selected Areas in Communications}, vol.~32, no.~6, pp. 1065--1082,
  Jun. 2014.

\bibitem{Adh14}
A.~Adhikary \emph{et~al.}, ``Joint spatial division and multiplexing for
  mm-wave channels,'' \emph{IEEE Journal on Selected Areas in Communications},
  vol.~32, no.~6, pp. 1239--1255, June 2014.

\bibitem{Lib99}
J.~C. Liberti and T.~S. Rappaport, \emph{Smart antennas for wireless
  communications: {IS}-95 and third generation {CDMA} applications}.\hskip 1em
  plus 0.5em minus 0.4em\relax Englewood Cliffs, NJ: Prentice Hall, 1999.

\bibitem{Ertel}
R.~B. Ertel \emph{et~al.}, ``Overview of spatial channel models for antenna
  array communication systems,'' \emph{IEEE Personal Communications}, vol.~5,
  no.~1, pp. 10--22, Feb 1998.

\bibitem{Durgin98}
G.~Durgin and T.~S. Rappaport, ``Basic relationship between multipath angular
  spread and narrowband fading in wireless channels,'' \emph{Electronics
  Letters}, vol.~34, no.~25, pp. 2431--2432, Dec 1998.

\bibitem{Rap13:Access}
\text{T. S. Rappaport} \textit{et al.}, ``Millimeter wave mobile communications
  for \text{5G} cellular: It will work!'' \emph{IEEE Access}, vol.~1, pp.
  335--349, 2013.

\bibitem{Sun14:CM}
S.~Sun \emph{et~al.}, ``{MIMO} for millimeter-wave wireless communications:
  beamforming, spatial multiplexing, or both?'' \emph{IEEE Communications
  Magazine}, vol.~52, no.~12, pp. 110--121, Dec. 2014.

\bibitem{Heath16}
R.~W. Heath \emph{et~al.}, ``An overview of signal processing techniques for
  millimeter wave {MIMO} systems,'' \emph{IEEE Journal of Selected Topics in
  Signal Processing}, vol.~10, no.~3, pp. 436--453, April 2016.

\bibitem{Samimi16_TMTT}
M.~K. Samimi and T.~S. Rappaport, ``{3-D} millimeter-wave statistical channel
  model for {5G} wireless system design,'' \emph{IEEE Transactions on Microwave
  Theory and Techniques}, vol.~64, no.~7, pp. 2207--2225, July 2016.

\bibitem{Can05}
E.~J. Candes and T.~Tao, ``Decoding by linear programming,'' \emph{IEEE
  Transactions on Information Theory}, vol.~51, no.~12, pp. 4203--4215, Dec.
  2005.

\bibitem{Ber14}
D.~E. Berraki \emph{et~al.}, ``Application of compressive sensing in sparse
  spatial channel recovery for beamforming in mmwave outdoor systems,'' in
  \emph{2014 IEEE Wireless Communications and Networking Conference (WCNC)},
  April 2014, pp. 887--892.

\bibitem{Alk15}
A.~Alkhateeby \emph{et~al.}, ``Compressed sensing based multi-user millimeter
  wave systems: How many measurements are needed?'' in \emph{2015 IEEE
  International Conference on Acoustics, Speech and Signal Processing
  (ICASSP)}, April 2015, pp. 2909--2913.

\bibitem{Alk14}
A.~Alkhateeb \emph{et~al.}, ``Channel estimation and hybrid precoding for
  millimeter wave cellular systems,'' \emph{IEEE Journal of Selected Topics in
  Signal Processing}, vol.~8, no.~5, pp. 831--846, Oct 2014.

\bibitem{Destino15}
G.~Destino \emph{et~al.}, ``Leveraging sparsity into massive {MIMO} channel
  estimation with the {adaptive-LASSO},'' in \emph{2015 IEEE Global Conference
  on Signal and Information Processing (GlobalSIP)}, Dec. 2015, pp. 166--170.

\bibitem{Malla16}
S.~Malla and G.~Abreu, ``Channel estimation in millimeter wave {MIMO} systems:
  Sparsity enhancement via reweighting,'' in \emph{2016 International Symposium
  on Wireless Communication Systems (ISWCS)}, Sept. 2016, pp. 230--234.

\bibitem{Gao16}
Z.~Gao \emph{et~al.}, ``Channel estimation for mmwave massive {MIMO} based
  access and backhaul in ultra-dense network,'' in \emph{2016 IEEE
  International Conference on Communications (ICC)}, May 2016, pp. 1--6.

\bibitem{Sun17_NYUSIM}
S.~Sun \emph{et~al.}, ``A novel millimeter-wave channel simulator and
  applications for {5G} wireless communications,'' in \emph{2017 IEEE
  International Conference on Communications (ICC)}, May 2017.

\bibitem{Aya14}
O.~E. Ayach \emph{et~al.}, ``Spatially sparse precoding in millimeter wave
  {MIMO} systems,'' \emph{IEEE Transactions on Wireless Communications},
  vol.~13, no.~3, pp. 1499--1513, March 2014.

\bibitem{Yu16}
X.~Yu \emph{et~al.}, ``Alternating minimization algorithms for hybrid precoding
  in millimeter wave {MIMO} systems,'' \emph{IEEE Journal of Selected Topics in
  Signal Processing}, vol.~10, no.~3, pp. 485--500, April 2016.

\bibitem{RapGut13}
T.~S. Rappaport \emph{et~al.}, ``Broadband millimeter-wave propagation
  measurements and models using adaptive-beam antennas for outdoor urban
  cellular communications,'' \emph{IEEE Transactions on Antennas and
  Propagation}, vol.~61, no.~4, pp. 1850--1859, Apr. 2013.

\bibitem{Rap15}
T.~S. Rappaport, R.~W. {Heath, Jr.}, R.~C. Daniels, and J.~N. Murdock,
  \emph{Millimeter Wave Wireless Communications}.\hskip 1em plus 0.5em minus
  0.4em\relax Pearson/Prentice Hall 2015.

\bibitem{Eka11}
C.~Ekanadham \emph{et~al.}, ``Recovery of sparse translation-invariant signals
  with continuous basis pursuit,'' \emph{IEEE Transactions on Signal
  Processing}, vol.~59, no.~10, pp. 4735--4744, Oct. 2011.

\bibitem{Mac17b}
G.~R. {MacCartney, Jr.} and T.~S. Rappaport, ``Study on {3GPP} rural macrocell
  path loss models for millimeter wave wireless communications,'' in \emph{2017
  IEEE International Conference on Communications (ICC)}, May 2017.

\end{thebibliography}

\end{document}